%% file: paper.tex
\documentclass[runningheads]{llncs}

\AtBeginDocument{%
  }

\usepackage{graphics} 
\usepackage[numbers]{natbib}
\usepackage{algorithmic}
\usepackage{graphicx}
\usepackage{enumitem}
\usepackage{caption}
\usepackage{diagbox}  
\usepackage{booktabs}
\usepackage{subfigure}
\usepackage{textcomp}
\usepackage{xcolor}
\usepackage{threeparttable}
\usepackage{multirow}
\usepackage{url}
\usepackage{tabularx}
\usepackage{array}
\usepackage{comment}
\usepackage{tabularray}

\newcommand{\M}{Item-Graph2vec}
\newcommand{\SG}{Skip-Gram}
\newcommand{\NV}{Node2vec}
\newcommand{\IV}{Item2vec}

\begin{document}
 
\title{\M: a Efficient and Effective Approach using Item Co-occurrence Graph Embedding for Collaborative Filtering}

\author{Ruilin Yuan\inst{1*}\and
Leya Li\inst{1*}\and
Yuanzhe Cai\inst{1}\\
\email{13510231378@163.com, lileya525@gmail.com, caiyuanzhe@sztu.edu.cn}}
\institute{Shenzhen Technology University, Shenzhen 518000, China 
\url{https://www.sztu.edu.cn/}}

\maketitle
\input{sections/Abstract.tex}
\input{sections/introduction.tex}
\input{sections/relatedWork.tex}

\input{sections/ItemGraph2vevAlgorithm.tex}

\input{sections/experimentalAnalysis.tex}

\input{sections/conclusion.tex}

\renewcommand{\bibname}{\leftline{References}} 
\bibliographystyle{splncs04}
\bibliography{references}

\end{document}

%% file: sections/Abstract.tex
\begin{abstract}
Current item-item collaborative filtering algorithms based on artificial neural network, such as \IV, have become ubiquitous and are widely applied in the modern recommender system.     
However, these approaches do not apply to the large-scale item-based recommendation system because of their extremely long training time. 
To overcome the shortcoming that current algorithms have high training time costs and poor stability when dealing with large-scale data sets, the item graph embedding algorithm \M \ is described here. 
This algorithm transforms the users' shopping list into a item co-occurrence graph, obtains item sequences through randomly travelling on this co-occurrence graph and finally trains item vectors through sequence samples. 
We posit that because of the stable size of item, the size and density of the item co-occurrence graph change slightly with the increase in the training corpus.
Therefore,~\M~has a stable run-time on the large-scale data set, and its performance advantage becomes more and more obvious with the growth of the training corpus.
Extensive experiments conducted on real-world data sets demonstrate that~\M~outperforms ~\IV \ by \textbf{3 times} in terms of efficiency on douban data set, while the error generated by the random walk sampling is small.

\keywords{Item embedding, Recommendation system, Co-occurrence graph, Random walk}
\end{abstract}

%% file: sections/introduction.tex
\section{Introduction}

The task of recommender systems is to generate a series of recommendation results that match user preferences using their historical behavior.
Item-item collaborative filtering, as a key building block in modern recommender systems, is a form of collaborative filtering based on the similarity score between items computed by people's clicks of those items. 
Item-item collaborative filtering was first published in an academic conference in 2001~\cite{sarwar2001item} and first applied by Amazon in 1998.

Recently, item similarity research works~\cite{barkan2016item2vec, barkan2020attentive, fang2020deep, martins2020deep} inspired by the deep learning technique in natural language processing (NLP) area achieves much better accuracy. 
These approaches apply the skip-gram with negative sampling (SGNS)~\cite{mikolov2013efficient} to collaborative filtering data to calculate the item embedding, since a sequence of words is equivalent to a set or basket of items. 
The similarity between items can be calculated by the $l_2$-distance or the cosine distance between two items' embedding.   
However, it is yet to be affordable to straightly apply these approaches to learning item embedding, because most of these methods suffer from a huge time consumption in training item embedding, especially for large-scale data sets. 

In this paper, we address the problem of learning item embedding on large-scale user-item clicking data sets. 
Our motivation comes from the word co-occurrence graph. 
Many researches~\cite{zuckerman2019using, sen2019word} in the NLP area have tried to map text to the word co-occurrence graph and capture the implicit structure of text data through the methods proposed by various graph models.
Inspired by these approaches, our solution is also established on a item co-occurrence graph, but different from the prior approaches, our approach do the item sampling by random walk traveling on this graph.
First,~\M~uses item co-occurrence information from a user's shopping list (or a user's movie watched list) to construct a item graph whose each node as a item, whose edges as co-occurrences between a item and its adjacency item in the list.
The item co-occurrence graph is the undirected graph because the relationship between the item and its adjacency item is bidirectional.
Second,~\M~performs fast and efficient random walk traveling on this item graph and samples item sequences. 
Third, the~\SG~model~\cite{mikolov2013efficient} has been applied to these sampled item sequences for learning the final item embedding.

We also argue that training the item embedding on a item co-occurrence graph is effective and efficient for the following reasons.
(i) \textbf{Effectiveness:} With random travelling on the item co-occurrence graph, our method can capture both \textbf{local} (in a local sliding window) and \textbf{global} (among different user' shopping list) item co-occurrence information together, and use this information to embed item under the assumption that highly co-occurring item should have similar meaning.
(ii) \textbf{Efficiency:} Two points motivate the improvement of our method's efficiency.  
First,~\IV~requires a long period to train the item embedding, since its training corpus of item clicking streaming is huge.
For example, amazon.com sold over 60,000 items per minute during the 2022 Prime Day and have 2.4 billion desktop and mobile visitors in May 2022. 
Therefore, it is huge time-consuming process to training the SGNS algorithm by sliding the window for scanning these user's clicking records.
Second, another interesting observation on item sequences is that there are a large number of same item co-occurrence segments among various users.
For example, if user A buy banana and milk, user B can also buy banana and milk again. 
The same shopping patterns are appeared again and again between various users, so that~\IV~wastes too much calculation time by sliding a window on these same shopping patterns.
Based on this observation, we emphasize that the item co-occurrence graph, which has already captured all the statistics of same shopping patterns, is a better model to analyze the item sequences. 
For example, as the number of nodes (items) is the relative small number (e.g., 62,423 movies from douban.com), these same watched movie patterns only change the edge weights of the item
co-occurrence graph, but the size of nodes and graph density do not have any alteration. 
Thus, it does not affect the~\M's training performance.

\subsection*{\textbf{Contributions:}}

\begin{itemize}[align=left,leftmargin=*]

    \item We first offer a detailed study on the item co-occurrence graph with consideration of different edge weights and nodes weights.   
	
    \item We present the~\M~approach, which is based on the item co-occurrence graph to calculate the item embedding. An effective sampling strategy is developed to generate the item sequences, and the~\SG~algorithm is used to calculate the item embedding on these item sequences (user's shopping list or movies watched list).  
	
    \item Extensive experiments are conducted to evaluate the efficiency and the accuracy of the proposed algorithm. Experimental results show that the~\M~algorithm is capable of outperforming~\IV~\cite{barkan2016item2vec} by 3.05 times on Douban data with better accuracy (6.4\% higher in top 200).
    
\end{itemize}

\noindent \textbf{Road Map:} 
The rest of this paper is organized as follows. 
Section~\ref{sec:related_work} first review the classic work on word embedding and graph embedding. 
Section~\ref{sec:algorithm} explains the detailed study of~\M~model. 
Section~\ref{sec:experiment} shows the experimental settings and empirical research. Section~\ref{sec:conclusion} summarizes the paper.

%% file: sections/relatedWork.tex
\section{Related Work}
\label{sec:related_work}

We divide the work related to our research into two categories: \textbf{word embedding methods and graph embedding methods}. 

\subsubsection{Word Embedding Method:} 
Word2vec~\cite{mikolov2013efficient} is the most typical word embedding method, Word2vec is divided into CBOW model and~\SG~model. 
CBOW model predicts the surrounding context of a given input word, while the~\SG~model predicts the input word through its surrounding context. 
FastText is also a text classifier, and its architecture is similar to CBOW model~\cite{barkan2016item2vec}, except that FastText predicts labels while CBOW predicts intermediate words. 
Words should have different meanings in different contexts, but Word2vec's vector representation of the word is the same, and FastText does a great job of solving this problem by learning complex features of word use and how these complex uses change in different contexts.
Meanwhile, \IV, a generalization of word2vec, extends it beyond NLP to recommendation system area. 
The user's shopping list or movies watched list is applied as the input for the~\SG~model. 
However, when larger data sets are encountered, the time cost of training the model will increase significantly. 
Our~\M~can reflect its uniqueness. 
As the data set size grows, the training time of the model will not change much, and it can still remain within a stable interval, while ensuring the accuracy of the recommendation.

\subsubsection{Graph embedding Method: }
Incorporating machine learning techniques on large-scale networks requires the learning of low-dimensional representations (embedding) of nodes. 
Node2vec~\cite{grover2016node2vec} is a popular method used to achieve this goal, which facilitates efficient analysis and interpretation of large graphs.
Node2vec is a graph embedding method that comprehensively considers the field of Depth First Search (DFS) and Broad First Search (BFS). 
\M~is based on the random walk mode in Node2vec. 
It uses two hyper-parameters $p$ and $q$ to control the probability in the random walk. 
Parameter $p$ controls the probability of repeatedly visiting the node just visited, and $q$ controls whether the walk is outward or inward.
While preserving both local and global information and accessing different types of nodes as much as possible,~\NV~can learn the relationships between movies by treating each movie as a node in a graph, and the relationships between them (such as shared actors, directors, production companies, etc.) as edges in the graph. By performing random walks on the graph.
\NV~can capture the relationships between movies, even if there is no direct connection between them, but through shared actors, directors, and other factors. 
After learning, the embedding vectors of movie nodes can be used to calculate the similarity between movies, enabling tasks such as movie recommendation.

To achieve efficient and high-quality node embedding for networks of all sizes and densities, we will use PecanPy~\cite{liu2021pecanpy} - a Python implementation of~\NV~which is fast, parallel, memory efficient, and cache optimized. 
PecanPy utilizes cache-optimized compact graph data structures and pre-computation techniques to provide optimal performance for graph embedding.

%% file: sections/ItemGraph2vevAlgorithm.tex
\section{\M~Algorithm}
\label{sec:algorithm}

\subsection{Motivation} 

Our motivation comes from the word co-occurrence graph, which have been successfully used in the NLP tasks, such as information retrieval~\cite{blanco2012graph} and text classification~\cite{hassan2007random}. 
These works use a word co-occurrence graph representation that captures these word relationships instead of the sliding window for traditional word embedding approach~\cite{barkan2016item2vec}. 
Similarly, we can extract item co-occurrence information from the item sequence (e.g., user's shopping sequence or user's movie watched sequence) to construct a item co-occurrence graph whose each item as a
node, whose edges represent co-occurrences between movies, and whose edge is undirected because the order of item should not be considered in the recommendation system.

\subsection{Framework of~\M}

\begin{figure*}[t]
    \centering
    \includegraphics[width=1\textwidth,height=30cm,keepaspectratio]{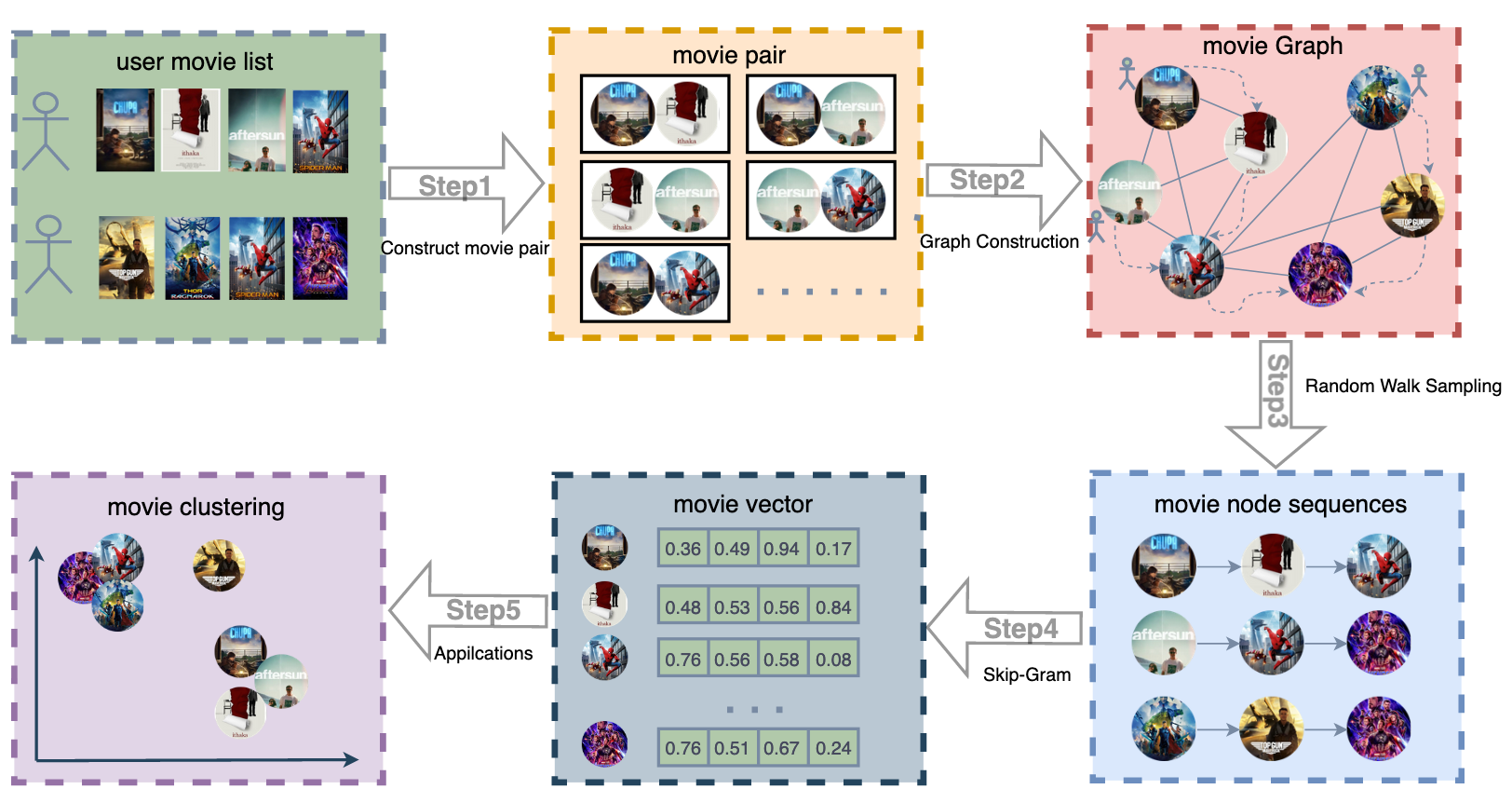}
    \captionsetup{labelfont={normalfont}}
    \caption{The framework of~\M~algorithm for movie embedding. 
    Step 1: The algorithm converts the user's movie sequences into the movies pairs to show their co-occurrence relationship. 
    Step 2: Build a movie co-occurrence graph based on movie pairs. 
    Step 3: Randomly walk on the movie graph to generate a series of movie sequences. 
    Step 4: Apply these movie sequences to the Skip-Gram approach to generate movie embedding. 
    Step 5: Apply these movie embedding for the recommendation system.}
    \label{fig1:word-graph2vec}
\end{figure*}

The Fig.~\ref{fig1:word-graph2vec} shows the framework of~\M, which is roughly divided into three steps. 
Take the user's movie watched list in MovieLens data set as an example. 
First, the user's movie watched list is used to construct the item co-occurrence graph. 
Second, the item sequence is generated from the item co-occurrence graph by random walking. 
Third, these sequences are used as the sample set, and the item embedding is trained by Skip-Gram algorithm~\cite{mikolov2013efficient}.

\subsubsection{(i)~Generate movie co-occurrence graph:} 

The user's watched movie list is rendered as a movie co-occurrence graph corresponding to a weighted graph whose nodes represent unique movies and whose edges represent co-occurrence between movies.

\begin{figure}
\centering
\subfigure[Users' movie watched list]{
\label{fig:movie-list}
\begin{minipage}[b]{0.7\textwidth}
\includegraphics[width=1\textwidth]{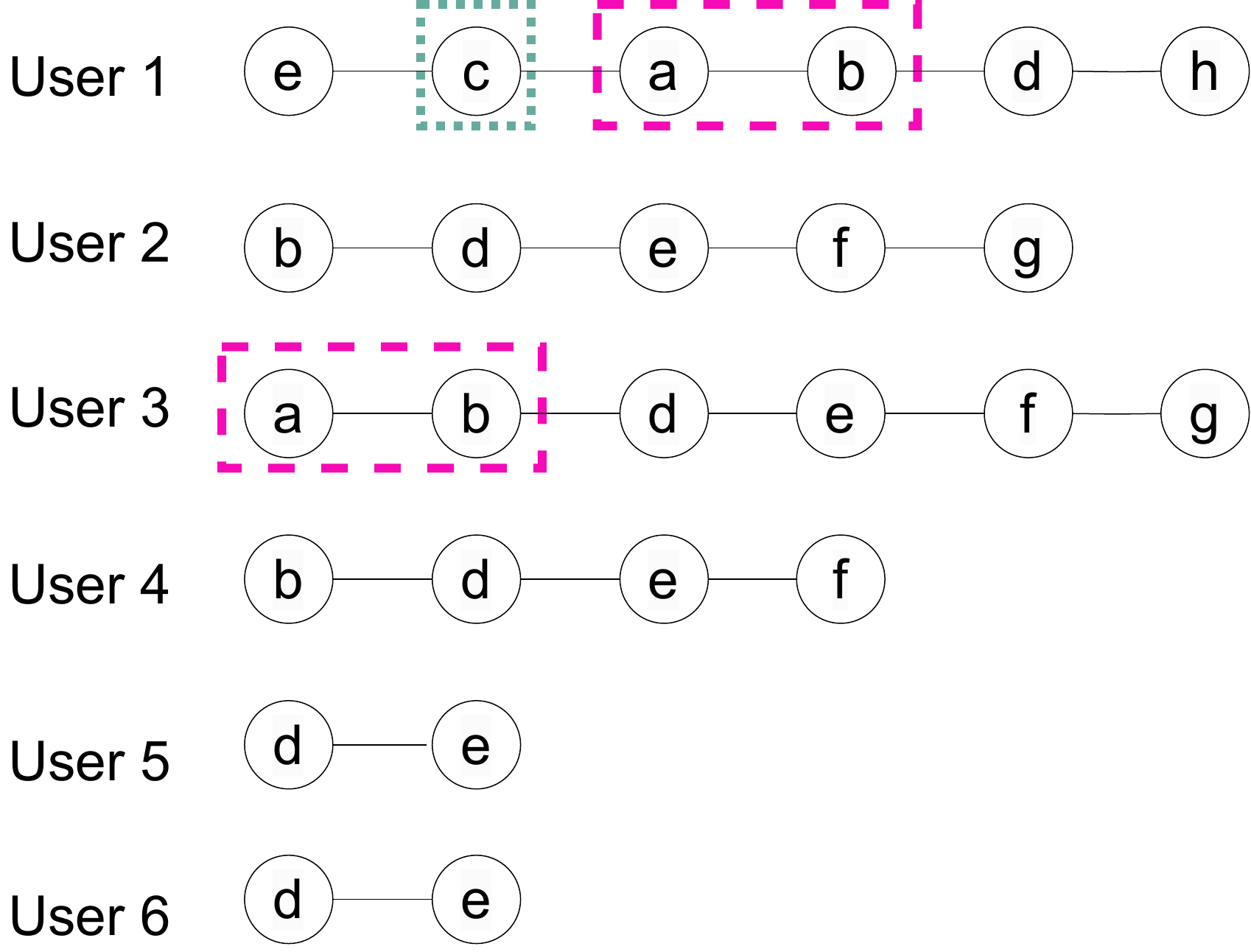}
\end{minipage}
}
\subfigure[A movie co-occurrence graph]{
\label{fig:movie-graph}
\begin{minipage}[b]{1\textwidth}
\includegraphics[width=1\textwidth]{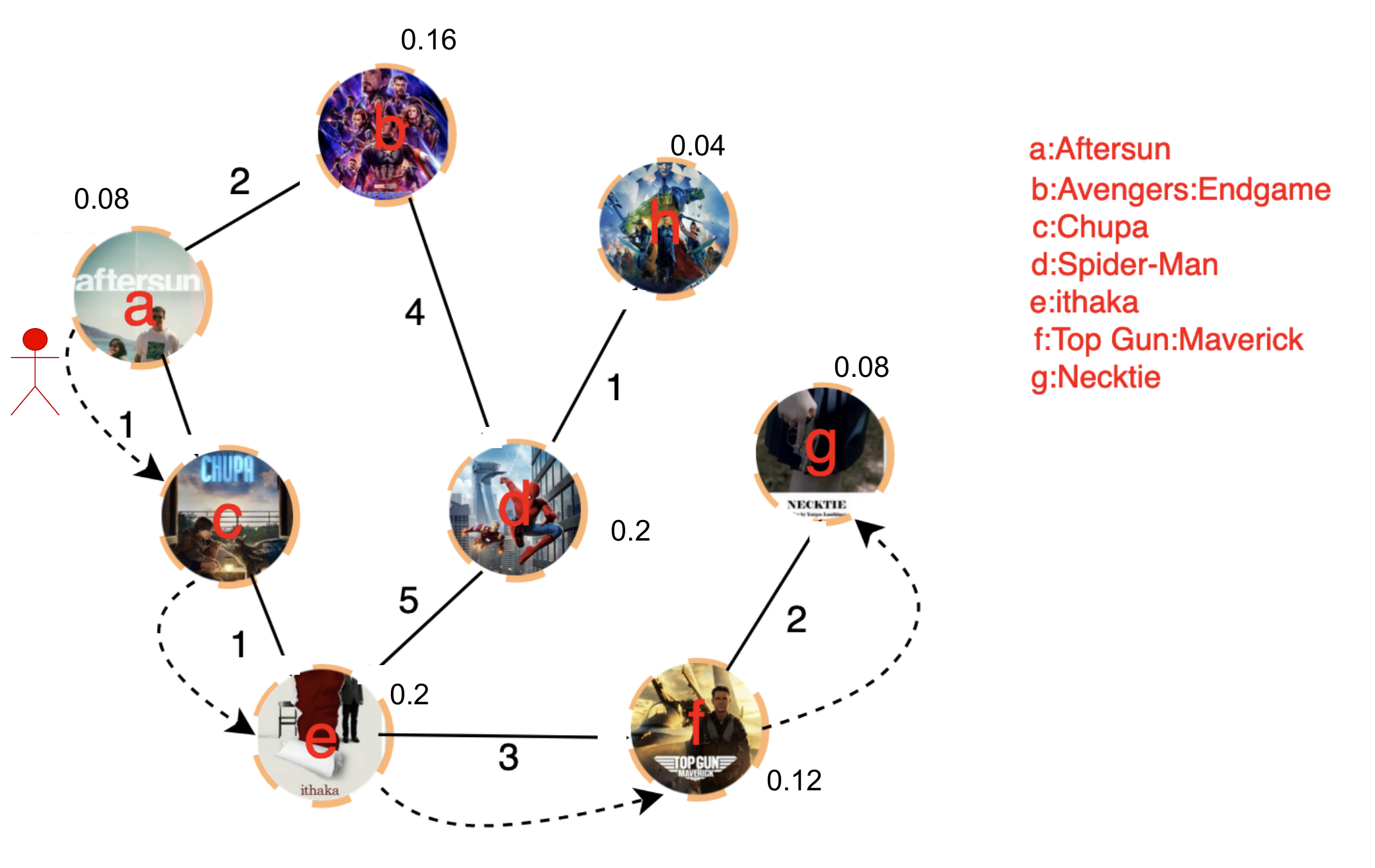}
\end{minipage}
}
\caption{An example of a movie co-occurrence graph. 
Fig~\ref{fig:movie-list} shows six users' watched lists. 
Fig~\ref{fig:movie-graph} shows the movie co-occurrence graph extracted from this users' watched lists. 
Each node represents a unique movie, and each edge is a co-occurrence of the two movies. 
} \label{fig:movie-graph-example}
\end{figure}

\textit{Weight on Edge:} 
The number of simultaneous occurrences of these two movies from the different users' are used as the weight of edges.
The weighted adjacency matrix $W$ is used to store the edges.
The weight, $W_{vx}$ is the weight from movie $v$ to movie $x$.

\begin{equation}
    W_{vx} = \sum co\mbox{-}occurrence(v,x)
    \label{equ:edge weight}
\end{equation}
where co-occurrences($v$, $x$) is the number of times that the movies $v$ and $x$ appear together from the different user's watched list. For example, in Fig~\ref{fig:movie-list}, the number of occurrences of the movie pairs of $a$ (Aftersun) and $b$ (Avengers Endgame) is 2, so the number of $W_{a,b}$ equals 2. 
Again, we get the relationship between each movie in the entire graph, and the movie co-occurrence graph is shown in Fig.~\ref{fig:movie-graph}. 

\textit{Weight on Node:} 
The weights of the nodes are used to determine the sampling frequency during the random walk. 
We take the probability $PW_v$ as the expression of the importance of the movie $v$ in the whole corpus, namely, the weight of nodes in the graph. 
Intuitively, the movie frequency can be applied to measure importance scores.
\begin{equation}
    PW_{v} = \frac{n_{v}}{\sum_{k} n_{k}}
    \label{equ:node weight}
\end{equation}
where $PW_{v}$ is the important score for movie $v$, $n_{v}$ is the number of occurrences of the movie $v$, and $\sum_{k} n_{k}$ is the number of occurrences of all movies. 
Taking Fig.~\ref{fig:movie-graph} as an example, movie \textbf{\textit{c}} appears 1 time and the occurrence number of total movies is 25, therefore $PW_{c} = \frac{1}{25} = 0.04$.

\subsubsection{(ii)~Sampling movie sequence by randomly walking:}

To estimate the closing degree between two nodes in a global structured network, we employed a random walk sampling procedure that iteratively explores the network. 
This approach enables us to gain insights into the structure of the network and capture the complex relationships between nodes. 
A random walk is generated from the current node, and random neighbors of the current node are selected as candidate nodes according to the transition probability on the movie co-occurrence graph. 
For example, in Fig.~\ref{fig:movie-graph}, one random walking sampling sequence is ``a $\rightarrow$ c $\rightarrow$ e $\rightarrow$ f $\rightarrow$ g", which is starting from movie a to movie g. 
According to the random walk travelling, The probability to surf from movie a to movie b is equal to 0.66 ($\frac{2}{3}$).  
In addition, these random walk sampling movie sequences have been collected as the training corpus for learning the item embedding.

Meanwhile, to better estimate the vector proximity, we carefully design the random walk procedure with the following \textit{two} characteristics.
First, the different nodes, as distinct from the same selection probability for graph embedding, should have different probabilities to be selected as the starting point. 
In other words, the probability of important nodes as the rooted node should be higher than the probability of unimportant nodes.
Intuitively, the probability sampling method based on
the importance score (see Equ.~\ref{equ:node weight}) is applied to select the rooted node.
For example, the sampling times of node $v$ as the starting node are shown in Equ.~\ref{equ:walk_nums}:
\begin{equation}
	number\_walks(v) = \left \lfloor total\_walks \times PW_{v} \right \rfloor
	\label{equ:walk_nums}
\end{equation} 
where total walks is the total number of random walk sampling \citep{perozzi2014deepwalk}. 
For example, let total walk = 200, number walks from movie $c$ should be 200 $\times$ $\frac{1}{25}$ = 8. 
It means there will be 8 movie sequences which is starting from movie ``c" (Chupa).

Second, to capture both homophily and structural proximity between movie nodes of the graph, we employ~\NV~second-order sampling strategy in~\M. In addition, Pecanpy which offers a parallelized and accelerated Node2vec in python, is applied in our experiments.

\subsubsection{(iii)~Training the item embedding with Skip-Gram model:}

For training movie embedding, we utilized the~\SG~model on sampled movie lists composed of a lot of random walks, which can be considered as sequences of movie nodes. 
The~\SG~model increases the \textit{co-occurrence probability} among movies that appear within a local context in a movie node sequence. As a result, the~\SG~model produces highly semblable vector representations for two movies that frequently appear within the local context of sampled movie node sequences.
Our sampling tactics significantly increases the possibility that these two movies are also frequently paired within local contexts in movie sequences, and so the distance ($l_{2}$-distance or cosine-distance) between embedding vectors of these two movies should be much small.

\subsection{Time Complexity Analysis}

We aim to address the issue of run-time efficiency in training movie embedding with the~\SG~model by comparing~\M~and~\IV. 
The time complexity of~\IV's training process is $O(N log(V))$, where $N$ is the total sampling movie library size and $V$ is the number of unique movie. 
On large data sets, where $N$ is large, training with~\IV~is computationally expensive. 
In contrast,~\M~utilizes a sampling movie library for~\SG~training, whose size is $nl$, where $n$ is the number of random walks and $l$ is the walk length. As a result, the time complexity of~\M~is reduced to $O(nl log(V))$, which is more efficient than~\IV~as $nl$ is smaller than $N$. Moreover, since $V$ is fixed and $nl$ is set by the user, our approach's training time remains quite \textbf{stable} for different sizes of data sets.

%% file: sections/experimentalAnalysis.tex
\section{Experimental Results}
\label{sec:experiment}

In this section, we provide an empirical evaluation of the~\M~method. We present clustering, precision test and performance test to evaluate the effectiveness of~\M~compared to the item-based algorithm,~\IV .

\subsection{Data sets}

We used different data sets to estimate our approach (see Table~\ref{tab:data set}).  
Our experimental code can be downloaded from Github~\footnote{\url{https://github.com/cpu135/item-Graph2vec}}.

Douban~\footnote{\url{https://github.com/lujiaying/MovieTaster-Open/tree/master/datas}} data set: A 25M data set consists of the evaluation data of multiple movies from multiple users, movie metadata information and user attribute information. 
It contains 34,893 different movies, 2,712 users and 1,278,401 co-occurrence relationships.

Movielens~\footnote{\url{https://grouplens.org/data sets/movielens/25m/}} data set: This data set contains the evaluation data of multiple movies from multiple users, movie metadata information and user attribute information. Up to 62,000 different movies, 162,000 users and 5,333,611 relationships are contained.

Anime~\footnote{\url{https://www.kaggle.com/data sets/hernan4444/anime-recommendation-database-2020}} data set: This data set includes a total of 26M user preference information of 73,516 animations from 12,294 users, presenting  66,263,320 the co-occurrence relationship.

Synthetic data sets: We use $I_i U_j$ to represent the item-user relationship with $i$ items and $j$ users.
All of the relationship have been randomly distributed between users and items. 
With the aim of assessing the time performance of our algorithm, we created the different size of user-item data sets with different densities, 0.1$\%$ and 1$\%$.

\begin{table}[]
\caption {The statistic of data sets}
\label{tab:data set}
\center
 \begin{tabular}{|c|c|c|c|c|}
    \hline
     \textbf{Data set}& \textbf{\#User}& \textbf{\#Item}    &\textbf{\#Relationship}\\
    \hline
    Douban&  183,123 & 62,423& 80,765,383\\	  
    \hline	
    Movielens&	162,000 & 62,000 &  5,333,611\\
    \hline
    Anime&  12,294 & 73,516 & 66,263,320\\
    \hline 
    \end{tabular}
\end{table}

\subsection{Experimental Setting}
The experiments were conducted on a PC with a 1.60GHz Intel Core 5 Duo processor, 8GB of RAM, and all the algorithms were implemented in Python. 
The cache-optimize ($d$) compact graph data structure and pre-computation were used to improve the training performance. 
The SparseOTF pattern was chosen based on Pecanpy's difference of running modes of networks with different densities, which is suitable for large and sparse networks, since the movie co-occurrence graphs constructed by current data sets are relatively sparse.
Meanwhile, we also tested three sets of $(p,q)$ parameters in the Pecanpy approach: (1, 0.001), (1, 1), and (0.001, 1). 
We found that the combination of (1, 0.001) achieved the highest accuracy. 
The parameter setting of q=0.001 can control the random walk process to explore further parts of the graph, which helps control the walking tendency of DFS and obtain better item embedding.
Based on these results, we used the parameters of p=1, q=0.001, dimension=128, walk length=80, and window size=10 in all subsequent experiments. 
In addition, other parameters of Pecanpy is set as the default value.
 
\subsection{Experiments and Results}

\subsubsection{Clustering Results}

For the purpose of compressing the movie vectors in the generated model, we present the 2D embedding produced by t-SNE \citep{van2008visualizing}, for~\IV~and~\M, respectively. 
The data set of Douban is used here, in which the clustering categories are divided into 6 film genres (e.g., crime, love, drama, documentary, action, animation) according to a web retrieved genre metadata, and different categories will show different colors. 

\begin{figure}[htbp]
    \raggedleft
    \captionsetup{labelfont={normalfont}}
	\includegraphics[width=1\textwidth,height=22cm,keepaspectratio]{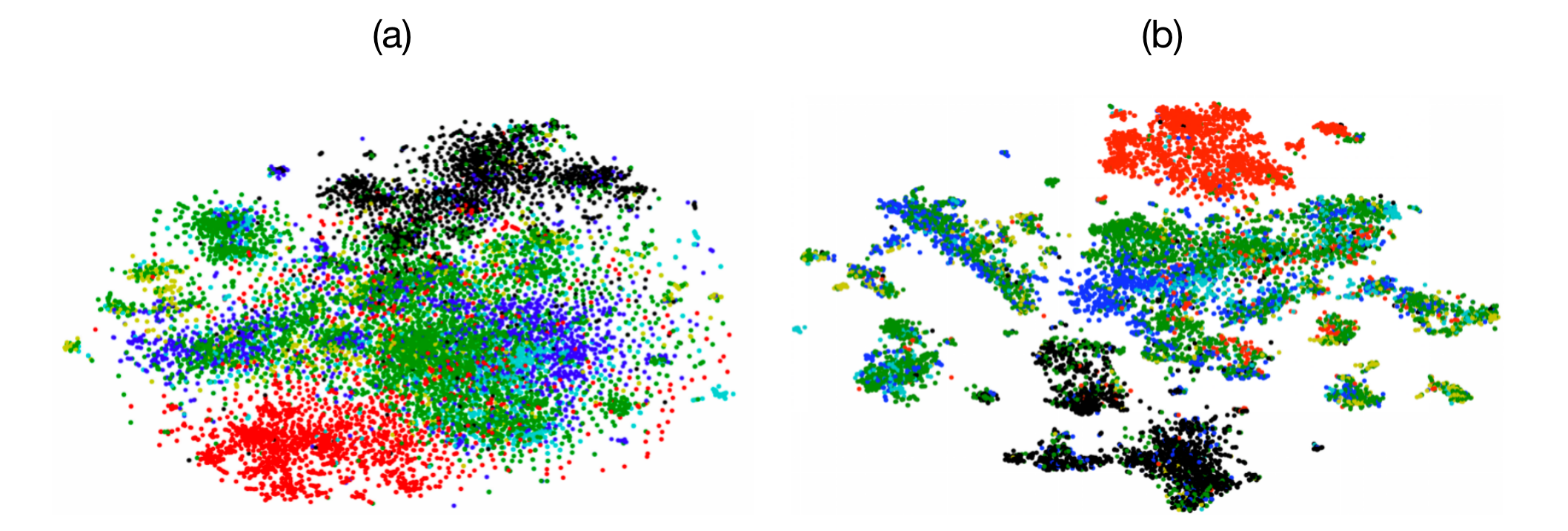} 
	\caption{T-SNE embedding for the item vectors produced by~\IV~(a) and~\M~(b). The items are colored according to six film genres (e.g., crime, love, drama, documentary, action and animation).
 }
	\label{fig:clustering}
\end{figure}

The clustering results of~\IV~and~\M~is shown in Fig.~\ref{fig:clustering}. It is evident that~\M~outperforms~\IV~in terms of clustering, as movie vectors with similar genres are more tightly clustered together in \M's embedding graph, while multiple movies overlapped in Fig.~\ref{fig:clustering}(a). 
This indicates that vectors generated by~\M~preserve the similarity information of movies while also having better discrimination. 
It is worth noting that even in the embedding graph, some similar movie vectors are still separated to some extent, which may be due to more complex similarity relationships in the actual data that require more data and sophisticated embedding algorithms to capture.

\subsubsection{Accuracy Experiment}

We define a accurate recommendation when the recommended movies have the same genre as the selected movie. 
This Experiment is mainly divided into two tasks. 
For the first one, we performed a accuracy comparison of the top 1-20 most similar movies in three data sets. 
Afterward, the algorithm accuracy comparison of top-10, 50 100 and 200 movies respectively on three data sets was completed. 
These experiments were conducted by searching for several movies which are most similar to a randomly sampled movie based on vector similarity.
We repeated the experiments 10 times and calculated the average results. 

\begin{figure}[t]
    \centering
    \subfigure[Douban data set]{
	\includegraphics[width=.45\linewidth]{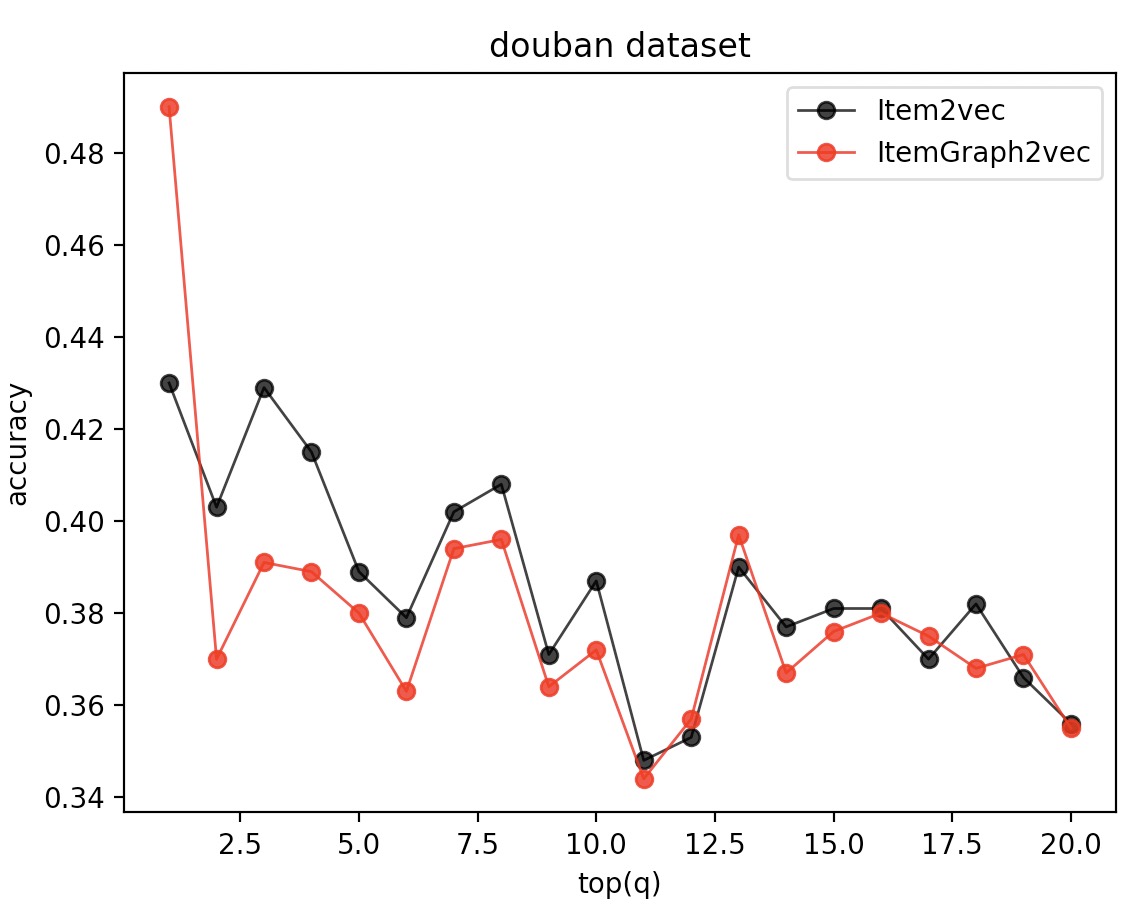} 
    \label{Fig:douban}}
    \hspace{0.1\linewidth}
    \subfigure[Movielens data set]
    {\label{Fig:movielens}
    \includegraphics[width=.45\linewidth]{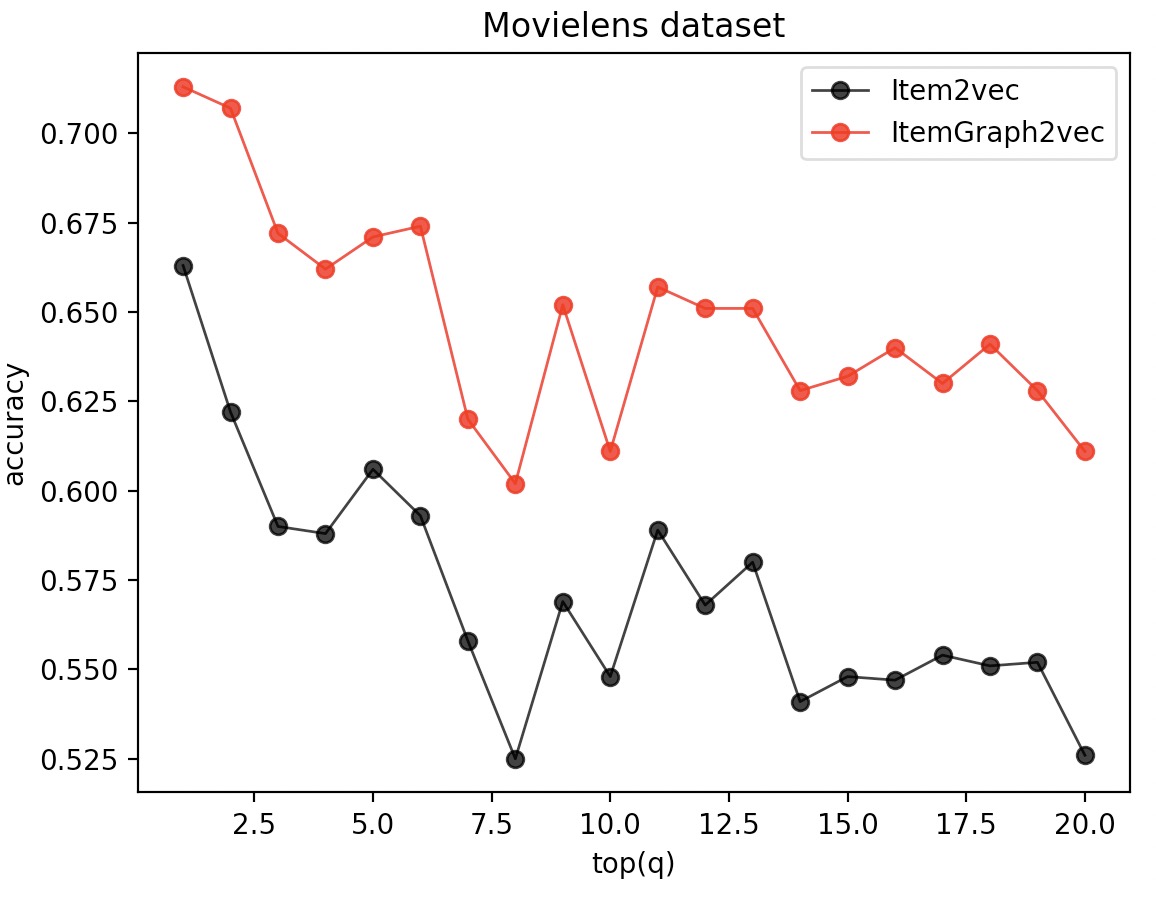}
    }
    \subfigure[Anime data set]{
	\includegraphics[width=.45\linewidth]{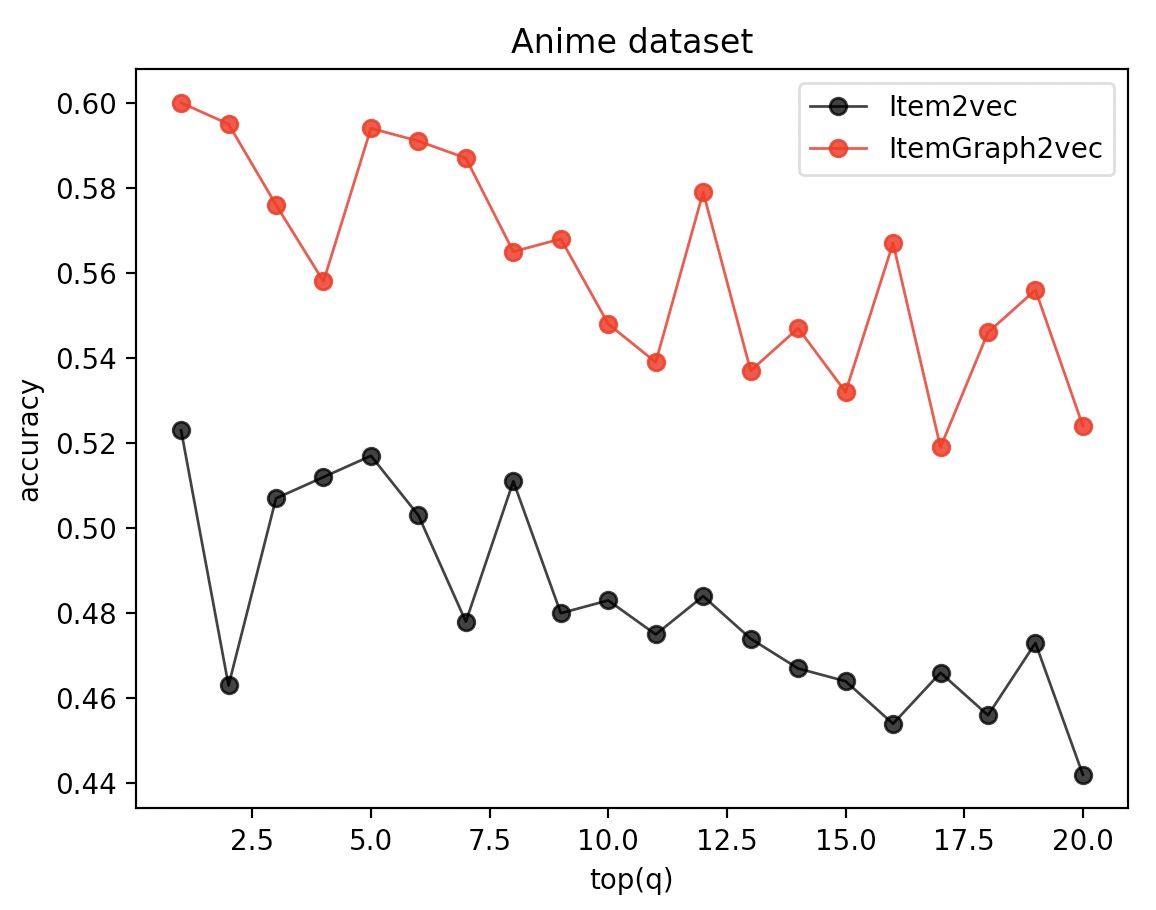} 
    \label{Fig:Anime}}
    \hspace{0.1\linewidth}
    \caption{\IV\ and \M\ precision comparison of the top 1 to top 20 most similar films on the Douban~(a), Movielens~(b), and Anime~(c) data sets.}
    \label{fig:top10}
\end{figure}

Fig.~\ref{fig:top10} compares the accuracy of the top 1-20 most similar movies. 
In general, \M~algorithm outperforms the \IV~algorithm, especially on Movielens data set and Anime data sets. 
The accuracy of \M~trained on these two data sets is about 10$\%$ higher than the other.
Differently, their accuracy are similar but \M~still performs slightly more precise on Douban data set. 
That indicates that the \M~algorithm performs better than the  \IV~algorithm in terms of accuracy. 

Table~\ref{tab:accuracy} describes the accuracy comparison of top-10, 50, 100, and 200 recommended movies on the three data sets. It can be concluded that \M\ recommends more precisely than~\IV~on most data sets in our experiment. 
\M~provides specific number of movies much more precisely for a higher value of around 10$\%$ on Movielens data set. 
\M~on Anime behaves slightly not that excellent compared to its performance on Movielens data set relatively but it remains a evident gap between the two methods.

\begin{table}
\centering
\caption{Comparison of algorithms accuracy of top-10, 50, 100 and 200 on three data sets.}
\label{tab:accuracy}
 \begin{tabular}{|c|c|c|c|c|c|c|}
\hline
\textbf{Accuracy}  & \multicolumn{3}{c|}{\textbf{\IV}} & \multicolumn{3}{c|}{\textbf{\M}}       
\\ \hline
Top-k & Douban   & Movielens & Anime   & Douban        & Movielens & Anime   \\ \hline
10     &\textbf{38.10}\%  & 57.70\%   & 48.20\% & 37.57\%       & \textbf{64.70}\%   & \textbf{55.73}\% \\ \hline
50     &\textbf{37.51}\%  & 50.91\%   & 41.53\% & 37.27\%       & \textbf{60.17}\%   & \textbf{50.54}\% \\ \hline
100    & 33.33\%  & 52.59\%   & 40.31\% & \textbf{34.23}\%       & \textbf{61.97}\%   & \textbf{47.66}\% \\ \hline
200    & 31.41\%  & 48.46\%   & 42.38\% & \textbf{33.45}\%       & \textbf{56.65}\%   & \textbf{48.44}\% \\ \hline
\end{tabular}
\end{table}

\subsubsection{Performance Experiment}

We conducted multiple experiments to evaluate the performance of~\M~algorithm in the domain of movie recommendation. 
We focus on time consumption and executed tasks from two aspects. 
On one hand, the time required for our algorithm to construct model generation vector was compared with~\IV, using all data sets. On the other hand, artificial data sets with different densities were applied to compare the influence of users and movies on model generation time. Densities of synthetic data sets are set at 0.1$\%$, 1$\%$, respectively, and the number of users and movies are set from 10,000 to 50,000.

Comparing the time consuming on training embedding on the all data sets, \M~algorithm is more stable and faster than~\IV, as shown in Table~\ref{tab:accuracy}. It took averagely 3 times less time for~\M~to generate vectors, especially when the models are trained on Anime, \M~saved about 1,400 seconds, \textbf{3.74} times shorter than~\IV. Among these three data sets, time consuming of~\IV~changed dramatically for at least 3,000 seconds, while time required by~\M~fluctuates 1,273 seconds at most. 

\begin{figure*}[h]
    \centering
    \captionsetup{labelfont={normalfont}}
	\includegraphics[width=1.0\textwidth,scale=1]{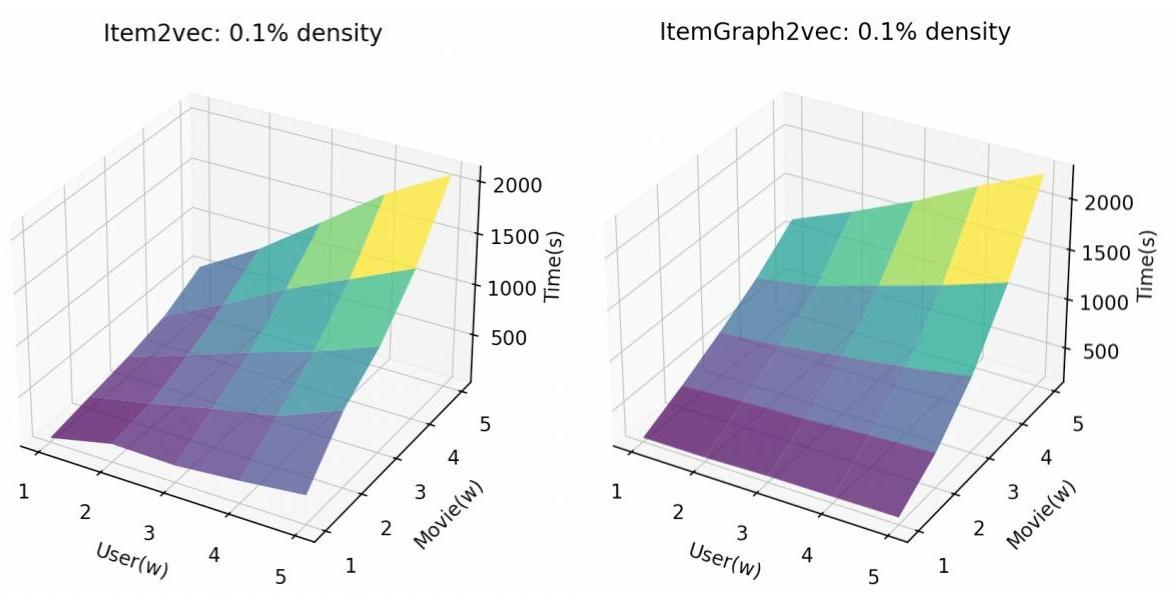}
    \label{Fig:density_0.1}
    \centering
    \captionsetup{labelfont={normalfont}}
	\includegraphics[width=1.0\textwidth,scale=1]{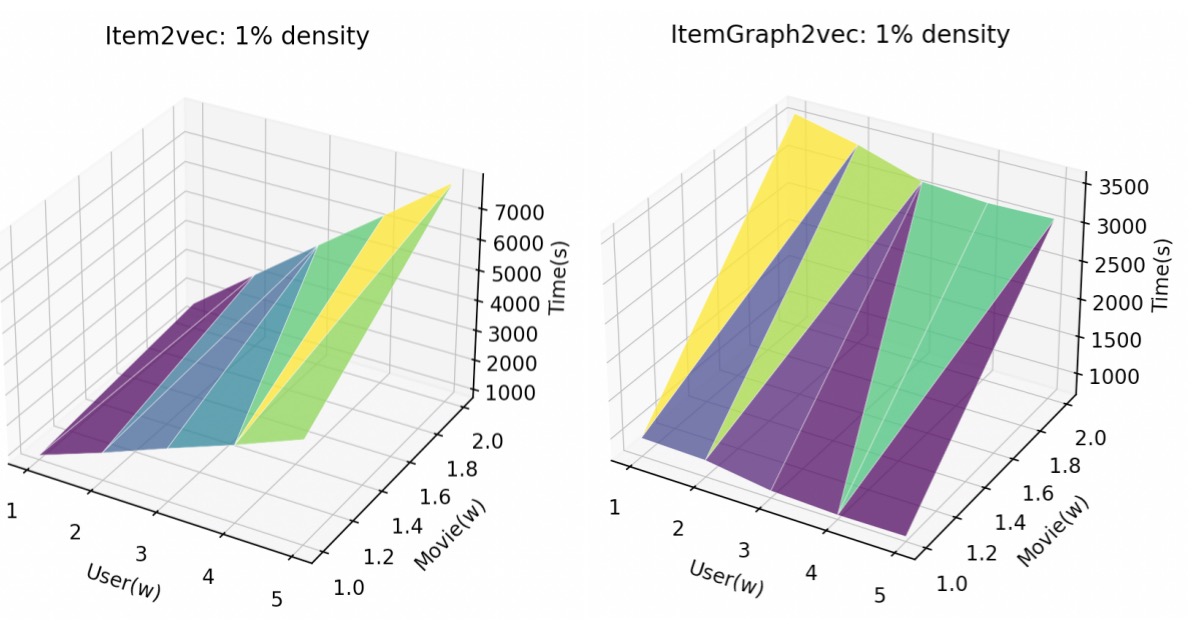} 
	\caption{The comparison of training time between~\M and~\IV on the different scale of synthetic data sets with densities at 0.1$\%$ and 1$\%$.} 
    \label{Fig:density_1}
\end{figure*}

Then, we compared the training time of \M \ and Item2vec on synthetic data sets with different densities, the different number of users and the different number of movies. 
The experimental results shown in Fig.~\ref{Fig:density_1} indicate: 
(i) Generally speaking,~\M~is a time-saving strategy rather than~\IV. 
For~\IV, increasing the number of users while keeping the number of movies constant results in a linear rise in overall time taken. 
By contrast, training time of~\M~is not greatly affected by the increasing number of users when the number of movies (nodes) remains the same. 
The number of movies remaining constant indicates that the number of nodes remains constant and growth in user only changes the edge weights maintaining the overall graph size. 
The running time of~\M~also increased linearly followed by an increase in the number of movies, but it is still more efficient than~\IV. 
Moreover, the larger the data set, the more obvious~\M's advantage in time performance is. 
When the density was 1$\%$, both of algorithms property fell rapidly compared to a lower density, while~\IV~performed seriously worse. 
At this density, only with a smaller number of users and a larger arrange of movies, did it take evidently more time for~\M~to represent relationships. 
On the whole, we confirm that~\M~performs better in generation speed, which benefits from its better movies accuracy.

\begin{table}
\centering
\caption{The training time between~\M~and~\IV~on various data sets.}
\begin{tblr}{
  cells = {c},
  hlines,
  vlines,
  hline{1,5} = {-}{0.08em},
}
\textbf{Data set} & \textbf{\IV} & \textbf{\M }\\
Douban data set& 20,784s    & \textbf{6,814s}  \\
Movielens data set& 15,224s    & \textbf{5,541s} \\
Anime data set & 18,994s    & \textbf{5,079s}          
\end{tblr} 
\end{table}

%% file: sections/conclusion.tex
\section{Conclusion}
\label{sec:conclusion}

In this paper, we proposed \M~an algorithm base on item co-occurrence graph for collaborative filtering. 
We compared the performance of \IV~and \M~on three tasks, including clustering, accuracy testing, and time performance experiments. The experimental results showed that~\M~generally had higher accuracy on the three data sets compared to~\IV, without losing accuracy due to the sampling method's characteristics, and performed exceptionally well in accuracy testing. 
For small data sets, the time consumption of~\M~and~\IV~may not differ much, but as the data set size grows,~\M's performance advantage becomes increasingly evident, with~\M~being able to outperform~\IV~by 3-4 times while maintaining a stable running time. This indicates that~\M~has excellent application prospects.